\documentclass[a4paper,twoside]{article}
\newcommand{\affil}[1]{$^{\rm #1}$}
\usepackage[authoryear]{natbib}
\bibpunct{(}{)}{;}{a}{}{,}
\usepackage{graphicx}
\date{} 
\def\arcdeg{\hbox{$^\circ$}}
\def\arcmin{\hbox{$^\prime$}}
\def\arcsec{\hbox{$^{\prime\prime}$}}
\def\ergcms{erg\,cm$^{-2}$s$^{-1}$}
\def\ha{H$\alpha$}
\def\hb{H$\beta$}

\def\nii{[N\,\textsc{ii}]}
\def\oiii{[O\,\textsc{iii}]}
\def\hii{H\,\textsc{ii}}

\def\p0{\phantom{0}}
\def\lessim{\raise-.5ex\hbox{$\buildrel<\over\sim$}}
\def\grtsim{\raise-.5ex\hbox{$\buildrel>\over\sim$}}

\title{\large\bf\flushleft PHL\,932: when is a planetary nebula not a planetary nebula? }
\author{\parbox{\textwidth}{\flushleft
\vspace{-0.5cm}
{\it David J. Frew\affil{A,B,E}, Greg J. Madsen\affil{C}, Simon J. O'Toole\affil{D}, and Quentin A. Parker\affil{A,D}}\\
\vspace{0.4cm}
{\small \affil{A}\,Department of Physics, Macquarie University, NSW 2109, Australia}\\
{\small \affil{B}\,Perth Observatory, Bickley, WA 6076, Australia }\\
{\small \affil{C}\,Sydney Institute for Astronomy, School of Physics, University of
  Sydney, NSW 2006, Australia}\\
{\small \affil{D}\,Anglo-Australian Observatory, PO Box 296, Epping
  NSW 1710, Australia}\\
{\small \affil{E}\,Email: dfrew@physics.mq.edu.au}}}
\begin{document}
\twocolumn[
\begin{changemargin}{.8cm}{.5cm}
\begin{minipage}{.9\textwidth}
\vspace{-1cm}
\maketitle
%
%
\small{\bf Abstract:}

The emission nebula around the subdwarf B (sdB) star PHL 932 is
currently classified as a planetary nebula (PN) in the literature.  Based
on a large body of multi-wavelength data, both new and previously published, we show
here that this low-excitation nebula is in fact 
a small Str\"omgren sphere (HII region) in the interstellar medium around this star.  We summarise the
properties of the nebula and its ionizing star, and discuss its
evolutionary status.  We find no compelling evidence for close binarity, arguing that PHL~932 is an ordinary sdB star.  We also find that the emission nebulae around the hot DO stars PG\,0108+101 and
PG\,0109+111 are also Str\"omgren spheres in the ISM, and along with
PHL~932, are probably associated with the same extensive region of
high-latitude molecular gas in Pisces-Pegasus. 



\medskip
\medskip
\end{minipage}
\end{changemargin}
]
\small

\section{Introduction}
\label{sec:intro}

\begin{figure}[h]
\begin{center}
\includegraphics[width=7cm]{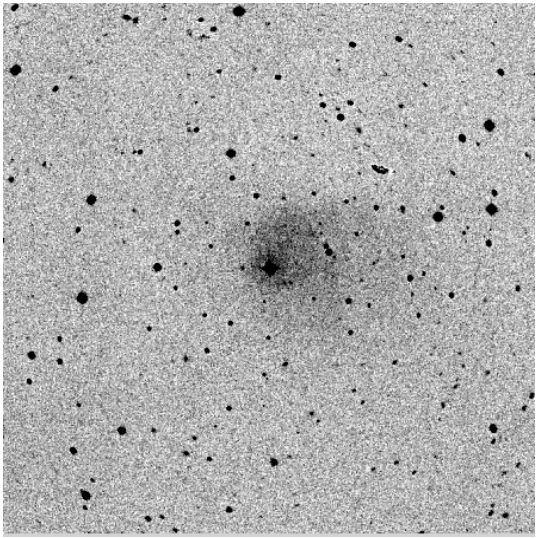}
\caption{The asymmetric emission nebula around PHL\,932.  The star is at the centre of this combined SuperCOSMOS broadband red and blue image, which is 10\arcmin\ across.}\label{fig:neb}
\end{center}
\end{figure}

PHL~932 was first noted as a 12th-magnitude blue star during a survey by Haro \&
Luyten (1962). It was studied in more detail by Arp \& Scargle (1967),
who discovered a faint emission nebula surrounding the star which they
assumed to be a planetary nebula (PN).  The nebula is 
asymmetrically placed around the ionizing star (Figure~\ref{fig:neb}), 
which has a position of $\alpha,\delta$ = 00$^{h}$59$^{m}$56.67$^{s}$, 
+15\arcdeg 44$^{m}$ 13.7$^{s}$ (J2000).  A spectrum of the nebula was
obtained by Arp \& Scargle (1967), who noted strong [O\,\textsc{ii}]
$\lambda$3727 and weak H$\alpha$ emission.  The almost complete absence of [O\,\textsc{iii}] emission (see \S\ref{sec:em_nebula} below) suggests the nebula has a very low ionization parameter, assuming it is photoionized.
An investigation of the stellar spectrum by M\'endez et al. (1988) placed the central star
squarely on the extreme horizontal branch (EHB) and allowed the
classification of PHL\,932 as a hot subdwarf~B (sdB) star.\footnote{M\'endez et al. (1988) classified the star as ``late sdO'' (or sdOB) based on their detection of HeII $\lambda$4686 in absorption.} 

Subdwarf B stars are unlike the typical central stars of PN.
They do not evolve to the asymptotic giant
branch (AGB), but rather move directly on to the white dwarf cooling curve.
Their progenitors are solar-like stars that have somehow shed
nearly all of their hydrogen envelopes prior to the onset of core
helium fusion (e.g. Heber 1986).
Exactly how this occurs is uncertain, although it is possible that some kind of binary
interaction is involved (Han et al. 2003). Indeed, De Marco et al. (2004) and Af\c{s}ar \& Bond (2005) have claimed the radial velocity of PHL\,932 is variable, which may indicate binarity (however, see Section~\ref{sec:binary}). More recently, De Marco (2009) has strongly argued for the general importance of binary interactions for the formation and shaping of PN. 

Despite the unusual nature of this system, the nebula around PHL~932
has not been studied in much detail since its discovery. 
In this paper, we present an in-depth analysis of the
nebula and the ionizing star based on new and archival
multi-wavelength observations. We use the results of this investigation to examine the
evolutionary status of the star and to reclassify the nebula.

\section{Observations}
\label{sec:obs}

This study makes use of new and previously unpublished
data, as well as already published observations. Details of the new
data are given here, while other archival observations used will be
cited as they are discussed.

\subsection{The Ionizing Star}

We use an archival ESO VLT/FORS1 spectrum of the star (Figure~\ref{fig:fit}), observed as part of an
investigation of magnetic fields in hot subdwarfs (Program
ID: 075.D-0352A). This spectrum was observed using the 600B grating in
spectropolarimetric mode with a slit width of 0.5\arcsec, which gives a
resolution of about 2.8\,\AA. The starlight passed through a Wollaston
prism and a rotatable retarder plate rotated to two different angles,
giving extraodinary ($e$) and ordinary ($o$) beams. Bias frames, flat
fields and He+HgCd arc spectra were taken at the end of the night of
observation. For the purposes of this study, the resulting $e$- and
$o$-spectra were combined, creating a Stokes $I$ spectrum with very
high signal-to-noise ratio. The data reduction was done using standard
routines in the \texttt{IRAF} package and the two spectra were reduced
separately and combined only after wavelength calibration.

\subsection{The Emission Nebula}

The large angular size and low surface brightness of the nebula
requires special instrumentation to measure accurately the emission
line properties. New observations of the nebula were obtained in 2005 with the
Wisconsin H-Alpha Mapper (WHAM) at the Kitt Peak National Observatory.
WHAM is a remotely operated Fabry-Per{\`o}t spectrograph designed to
detect very low-surface brightness optical emission lines at high
spectral resolution of R~= 25 000 (Reynolds et al. 1998).

WHAM records the spatially averaged spectrum over a 60\arcmin\ circular
diameter beam.  Figure~\ref{fig:WHAMneb} shows WHAM spectra of the H$\alpha$,
[N\,\textsc{ii}] $\lambda$6584 and [O\,\textsc{iii}] $\lambda$5007
lines toward PHL\,932.  Contamination from atmospheric lines were
removed, and the spectra were calibrated using synoptic observations
of NGC\,7000 (tied to results from the WHAM Northern Sky Survey of
Haffner et al. 2003) as well as a range of PN calibrators.  The spectra are displayed in units of
milli-Rayleigh per kms$^{-1}$; the secondary peak at $v_{\mathrm{LSR}} = -30$
km~s$^{-1}$ in the H$\alpha$ spectrum is an artefact of imperfect subtraction of
the bright geocoronal H$\alpha$ line.  

In Sep 2008 we observed the nebula with the SPIRAL integral field unit (Sharp 2006) on the 3.9-m AAT, located at Siding Spring Observatory, NSW.   The SPIRAL IFU is a 22.4\arcsec\ $\times$ 11.2\arcsec\ 512-element lenslet array linked by fibre feed to the AAOmega spectrograph. It is placed at the Cassegrain focus, uses the f/8 secondary and provides a spatial resolution of 0.7\arcsec. Standard 685V and 385R medium and low resolution gratings were employed in the blue and red arms of the AAOmega spectrograph.

\section{Evolutionary status of \\ PHL\,932}
\label{sec:evol}

\subsection{Spectral analysis and distance}
\label{sec:spec}
\begin{figure}[h]
\begin{center}
\includegraphics[width=7.5cm]{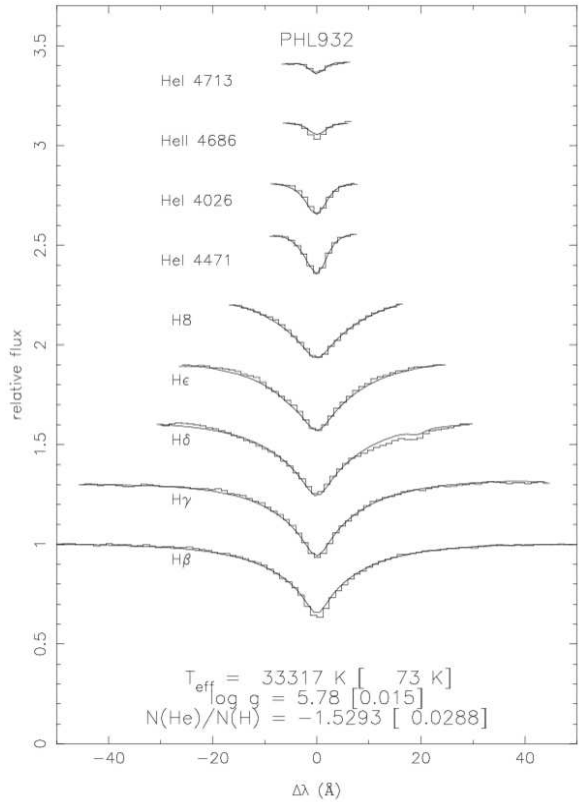}
\caption{Model fit to the individual lines in the VLT FORS1 spectrum of PHL\,932. The spectrum is
  represented by a histogram, while the best-fit model is given as an overlaid
  solid line.}
\label{fig:fit}
\end{center}
\end{figure}

We have used the archival, high signal-to-noise FORS1 spectrum to derive the
stellar parameters of PHL\,932 using the same method and model grid
([Fe/H]=0.0) as Lisker et al. (2005). We find $T_{\mathrm{eff}} =
33\,490\pm73$\,K, $\log\,g = 5.81\pm0.02$ and log(He/H)$ =
-1.58\pm0.03$, or in other words, consistent with being a
typical hot sdB star. The model fit to selected absorption lines is shown in Figure
\ref{fig:fit}. We note that the line profiles of both the
He\,\textsc{i} 4471\,\AA\ and He\,\textsc{ii} 4686\,\AA\ lines are
well matched by the synthetic spectrum. In several sdB stars with
similar parameters, there is a mismatch between the two ionization
species (Heber et al. 2000). O'Toole \& Heber (2006) investigated this
further after discovering large over-abundances of iron-group elements
using high-resolution UV echelle observations and found that models
with enhanced metals ([Fe/H]=+1.0) could remove or reduce the
discrepancy. They also found that there was no discrepancy for the one
star they studied with reduced iron abundance (Feige~66).   It is
likely that PHL\,932 has a similarly depleted iron abundance; indeed,
this is suggested in the work of Edelmann (2003). Alternatively, there
may be a resolution effect clouding the issue. Further high-resolution
optical studies are needed to clarify the situation.

Inferring the evolutionary status of the ionizing star depends
critically on its distance.  It is only recently that the distance has
been determined accurately.  M\'endez et al. (1988) derived a
spectroscopic parallax (a distance derived using surface gravity, but
assuming a stellar mass) of $\sim$390~pc after assuming
$M \simeq\,0.3\,M_{\odot}$, while Napiwotzki (1999) found a distance of 235~pc using the same method.   

PHL\,932 was observed by the \emph{Hipparcos} satellite.  The revised \emph{Hipparcos} catalog gives a
distance of 113$^{+114}_{-38}$ pc (van Leeuwen 2007), however this value, as with the original \emph{Hipparcos} parallax leads to an non-physical stellar mass for PHL\,932 of $<$0.1\,$M_\odot$. An improved parallax of 3.36 $\pm$ 0.62 mas (Harris et al. 2007) has shown that
the Hipparcos value is in error.  We adopt the new parallax distance of
298$^{+67}_{-47}$\,pc hereafter.  Combining this distance with our new gravity determination above
leads to a possible mass range of 0.3 -- 0.6\,$M_\odot$, which is
consistent with the expected properties of a hot sdB star. This distance
does not constrain the evolutionary history of PHL\,932
however.

O'Toole (2008) discussed the possible evolution channels that
can produce a spectroscopically identified hot subdwarf and ways to
distinguish them. In brief, a star that suffers extreme mass-loss on the red giant branch may either
ignite helium or not, depending on the amount of mass lost. In the
former case, the star will become an EHB star and eventually become a
C/O white dwarf with a mass of $\sim$0.48\,$M_\odot$, while in the latter case,
the star will become a He-core white dwarf with mass $<$0.46\,$M_\odot$.
See also Brown et al. (2001) and Han et al. (2002, 2003) for further discussion
of potential evolutionary channels for hot subdwarf stars. Currently,
with a mass range for PHL\,932 which covers several
possibilities, it is not possible to determine the correct
channel. Only a more precise distance determination will allow this.

\subsection{Non-binary status of PHL\,932}
\label{sec:binary}

In a spectroscopic study investigating the binary status of PN central stars,
De Marco et al. (2004) claimed a 98\% probability that PHL\,932 is a radial velocity variable, based on one discrepant radial velocity measurement out of nine.  However, when we look at other radial velocity
measurements in the literature, we find that radial velocity variability is
unlikely.  Arp \& Scargle (1967) give $V_{\rm hel}$ = +15 $\pm$ 20
kms$^{-1}$ based on a low-dispersion blue spectrum. Using
high-resolution echelle spectroscopy, Edelmann (2003) found $V_{\rm
  hel}$ = +18 $\pm$ 2 kms$^{-1}$ ($V_{\rm LSR}$ = +16.4 $\pm$ 2\,kms$^{-1}$) with no evidence for radial velocity variability.\footnote{In the direction of PHL 932, $v_{\mathrm{LSR}}$ = $v_{\mathrm{helio}}$ $-$ 1.6 kms$^{-1}$}  M\'endez (1989) also used echelle spectroscopy and found an
identical value of $V_{\rm hel}$ = +18 $\pm$ 2\,kms$^{-1}$ based on two spectra.  Wade (2001) used spectra obtained with the Hobby-Eberly Telescope to also show that no large radial velocity amplitude exists.  

Furthermore, Bond \& Grauer (1987) photometrically observed the 
central star but found no evidence for any variation due to irradiation effects, making it unlikely to be a close binary.  In addition, the spectral energy distribution derived from available optical and 2MASS data (Arp \& Scargle 1967; Wesemael et al. 1992; Allard et al. 1994; Harris et al. 2007; Cutri et al. 2003) is 
consistent with no near-IR excess from a cool companion.  

Given the constancy of the high resolution RV data of Edelmann
(2003), supported by  M\'endez (1989), Wade (2001) and Saffer, Green
\& Bowers (quoted by Wade 2001), we believe it is unlikely that PHL\,932 is a {\it close} binary system.

\subsection{Kinematics}
\label{sec:kin}

While the Hipparcos parallax is in error, the catalogued proper motion
(van Leeuwen 2007) for PHL 932 agrees well with the value in the UCAC2 catalogue (Zacharias et al. 2004),  and the new determination by Harris et al. (2007).  We adopt the UCAC2 value of 36.6 mas yr$^{-1}$ in PA
83\arcdeg.  The tangential velocity at the adopted distance of 298~pc
is then 52 $\pm$ 10 kms$^{-1}$.

Using the mean heliocentric radial velocity of +18 $\pm$ 2 km~s$^{-1}$
(M\'endez 1989; Edelmann 2003) and following the precepts of Johnson
\& Soderblom (1987),  the space motion vectors are determined to be $U
= -52~$km~s$^{-1}$, $V = -15$~km~s$^{-1}$, $W = -7$~km~s$^{-1}$. 
These vectors are typical for the kinematics of old-disk stars.  
The total space motion is $\sim$55~km~s$^{-1}$ with respect to the Sun, or $\sim$42~km~s$^{-1}$
with respect to the local standard of rest.  The space motion is
consistent with  the kinematics of sdB stars, most of which belong to
the old disk (Thejll et al. 1997), though some halo objects are known.

\section{Nebular Properties}
\label{sec:em_nebula}

\begin{figure}[h]
\begin{center}
\includegraphics[width=7.5cm]{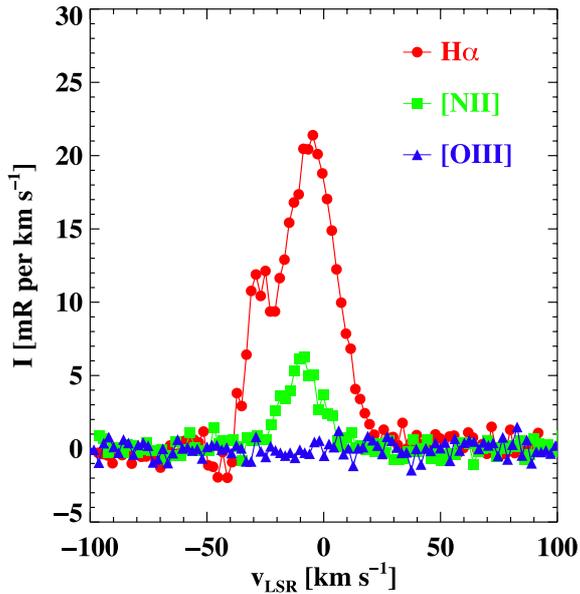}
\caption{WHAM spectra through a 60\arcmin\ beam centred on PHL\,932.}\label{fig:WHAMneb}
\end{center}
\end{figure}

We use the WHAM spectra (Figure~\ref{fig:WHAMneb}) to calculate the total \ha\ and \nii\ fluxes of the emission nebula, finding log\,F(\ha) = $-10.67$ $\pm$ 0.05\,\ergcms,  log\,F(\nii) = $-11.32$ $\pm$ 0.05 \ergcms, and \oiii\ emission too faint to be detected over the 60\arcmin\ WHAM beam.  
Note that the large WHAM aperture includes the extended ionized wake (see below).  The \ha\ nebula was also reported by Reynolds et al. (2005), using lower signal-to-noise WHAM data, and cataloged  as WPS\,37.  They found an integrated \ha\ flux of log\,F(H$\alpha$) = $-10.75$ $\pm$ 0.10 \ergcms, but this beam was not centred on PHL\,932.

To confirm the accuracy of the WHAM \ha\ flux, we performed aperture photometry on 
images from the Southern H-Alpha
Sky Survey Atlas (SHASSA; Gaustad et al. 2001), following the precepts
of Frew, Parker \& Russeil (2006).  The estimated flux is log\,F(\ha)
= $-10.73$ $\pm$ 0.06 \ergcms\ through a 12\arcmin\ aperture (mean of
two SHASSA fields, \#235 \& 236), after adopting {[\rm N \sc ii]}/\ha\
= 0.3  $\pm$ 0.1 from our WHAM data.  The measured \ha\ flux  through a larger
60\arcmin\ aperture is log\,F(\ha) = $-10.63$ $\pm$ 0.06 \ergcms\ (mean
of two fields), in excellent agreement with the measured integrated
flux from our new WHAM data.

There is a simple linear relationship between \ha\ surface brightness, an observational property, and emission measure ($\int$n$_e^2$d$l$), a physical property, provided a nebula is photoionized and has a temperature near T$_e$=10$^4$K (see Madsen, Reynolds \& Haffner 2006).  
The surface brightness of the emission nebula around PHL\,932 is non-uniform; we use the higher angular resolution SHASSA data to measure the peak surface brightness.  Using both SHASSA
fields, we find concordant results with a peak \ha\ surface brightness
of 36~Rayleighs, or $2.1 \times 10^{-16}$ \ergcms arcsec$^{-2}$ (corrected for background and {[\rm N \sc ii]} contribution).  This implies an emission measure of 80--100 cm$^{-6}$~pc, accounting for the uncertainties in surface brightness and nebular temperature.  The SHASSA data also reveal an extended 5 Rayleigh tail of emission that subtends ~18\arcmin\ southwest from the ionizing star; this is opposite to the direction of the proper motion vector and implies the tail may be some form of a wake.  

\begin{figure}[h]
\begin{center}
\includegraphics[width=5.5cm,angle=-90]{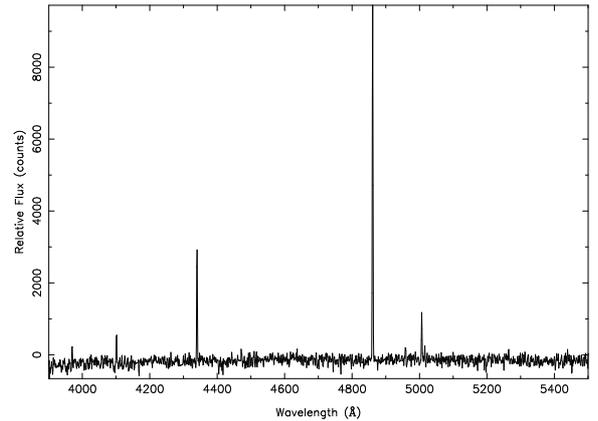}
\caption{Blue-arm, wavelength-calibrated AAT SPIRAL spectrum of the brightest part of the nebula, 20\arcsec\ NW of PHL~932. Note the very low nebular excitation, with an $\lambda$5007/\hb\ ratio of $\sim$0.1.}
\label{fig:SPIRAL_spec}
\end{center}
\end{figure}

We also comment on our SPIRAL data.   We decided to forego any spatially resolved information from the IFU and binned up all the spaxels to get maximum S/N in the spectrum of this faint nebula.  The binned, wavelength calibrated spectrum from the blue arm is shown in Figure~\ref{fig:SPIRAL_spec}.  The data are not flux calibrated, but we can extract useful line ratio data for emission lines close together in wavelength.  We measured a $\lambda$5007/\hb\ ratio of only $\sim$0.1, which is much lower than is typical for bona fide evolved PN of comparable surface brightness (Frew 2008).   This could be explained by the low temperature of the CS of only 33 000\,K.  While we note this very low level of excitation is sometimes seen in bona fide PN, these are all very young, highly compact nebulae that have cool central stars evolving to the left in the HR diagram.  None has a radius as large as the emission nebula surrounding PHL~932.

Preliminary inspection of the spatially resolved data shows that the \oiii\ emission is confined to an area closest to the ionizing star.  We also note a \nii/\ha\ ratio of 0.7 from the SPIRAL red-arm data, which is higher than the value determined from WHAM for the whole nebula (integrated over a 1\arcdeg\ beam).  This suggests the electron temperature of the gas is higher closer to the star.

The mass of the ionized nebula may be estimated from its distance, measured angular size, and the integrated \ha\ flux (following the recipe of Hua \& Kwok 1999).  Using the appropriate values for PHL\,932, we find the ionized mass is 0.04$\sqrt{\epsilon}$\,$M_{\odot}$, where $\epsilon$ is the (unknown) volume filling factor. This is an order of magnitude less than the median PN mass of 0.5$\sqrt{\epsilon}$\,$M_{\odot}$ (Frew \& Parker 2009).  

The radial velocity of the ionized gas is provided by the high spectral resolution WHAM data. A Gaussian fit to the line profiles in Figure~\ref{fig:WHAMneb} yield velocity centroids of $v_{\rm LSR}$ = $-5$ $\pm$ 3\,kms$^{-1}$ for \ha\ and $v_{\rm LSR} = -13 $ $\pm$ 3 kms$^{-1}$ for \nii; we adopt the mean value of $v_{\rm LSR}$ = $-9$ $\pm$ 4 kms$^{-1}$ as the velocity of the nebula.  The \texttt{IRAF} {\sl emsao} package was used to determine an emission-line velocity for the nebula from our SPIRAL data.  Using the higher-resolution blue data (excluding two weak lines), we derive  $v_{\rm LSR}$ = $-3$~$\pm$~5 kms$^{-1}$,  in excellent agreement with our WHAM results.

The velocity from WHAM can be compared to the stellar radial velocity of PHL~932 (M\'endez 1989; Edelmann 2003) of $v_{\rm LSR}$ = +16.4 $\pm$ 2 kms$^{-1}$ (see \S\ref{sec:binary}). The radial velocity of the nebula differs from the radial velocity of the ionizing star by $\sim$6$\sigma$. This is very strong evidence that the two are not associated, especially since the star does not exhibit radial velocity variability.

Lastly we use a Gaussian fit to the WHAM data to measure the line widths of the gas. We find an \ha\ full-width half-maximum line width of 2$v_{\rm exp}$ = 22 $\pm$ 4 kms$^{-1}$.  This line width is typical of diffuse interstellar ionized gas and significantly lower than the average width seen in evolved PN, $\sim$50 km s$^{-1}$ (Reynolds et al. 2005).

\section{Identity of the Nebula}

The nebula surrounding PHL\,932 has long been categorized as a PN (e.g. Acker et al. 1992).  
M\'endez et al. (1988) assessed the identity of the nebula and argue that their estimate of its emission measure ($\sim$60 cm$^{-6}$pc) is too large to be attributed to the ambient interstellar medium.
They conclude it must have been ejected from the star and affirm its status as a `planetary nebula', albeit a peculiar one with an sdB central star (see Frew \& Parker 2009, for a review of PN taxonomy).

We have presented new data that strongly suggest that the nebula is, in fact, an \hii\ region unrelated to the star.  Firstly, independent measures of the radial velocity of the gas are significantly different from the radial velocity of the star. Second, the nebular line width is typical of the ionized ISM and is lower than is seen in most evolved PN.  Third,  the ionized mass is significantly lower than is expected for an evolved PN.  Fourth, the \oiii/\hb\ ratio shows the level of excitation is very low, much less than for any evolved PN.   Lastly, the detailed nebular morphology is not consistent with the emission nebula being ejected from PHL~932.  Indeed, the morphology is irregular and the surface brightness essentially drops off away from the ionizing star in all directions.\footnote{Deep \ha\ and \nii\ images taken by C.-T. Hua with the OHP 1.2-metre reflector show a non-limb brightened, irregular nebulosity with apparent dust lanes.  The morphology is consistent with the nebula being a small HII region.  See http://www.oamp.fr/people/trung/phl932.html}  From section~\ref{sec:kin}, the total velocity with respect to the local standard of rest is $\sim$43 kms$^{-1}$.  Despite the moderately fast space motion, there is no sign of an enhanced rim or  bowshock on the leading side of the nebula (see Figure~\ref{fig:neb}) as expected from theoretical modelling (e.g. Wareing, Zijlstra \& O'Brien 2007) of PN/ISM interactions.  In the light of this, we find that the nebula is not moving through the ISM and we interpret the observed wake as a recombining fossil `contrail'.

In summary, we conclude that the star is in an overdense region of the ISM and has ionized a small volume around it.  To further assess this scenario, we consider the interstellar environment of PHL\,932.
There are at least three high-latitude molecular clouds in the broad vicinity,
part of a large arc of CO emission (Magnani et al. 2000) which may
link a south-extending spur from the Taurus-Auriga dark cloud complex
with the eastward extension of the Pegasus molecular cloud association (note that a recent detailed
study in CO  of the extended Pegasus region has been undertaken by
Chastain et al. 2006).   
The molecular cloud MBM~3 (Magnani, Blitz
\& Mundy 1985) is found 4\arcdeg\ northeast of PHL~932 and has a
similar systemic radial velocity to the emission nebula around PHL~932.  
This cloud seems to be associated with the diffuse nebula
LBN~639 (Lynds 1965), listed in SIMBAD as a HII region, but
is more likely to be a reflection nebulosity, or galactic cirrus, as
an enhancement is not present in the WHAM-NSS data (Haffner et
al. 2003).  Other molecular clouds are nearby; MBM~4 (associated with
LBN~644) is 2\arcdeg\ east of MBM~3 and linked to it, while
MBM~2 is 7.5\arcdeg\ southwest of PHL 932. 

\begin{figure}[h]
\begin{center}
\includegraphics[width=7.5cm]{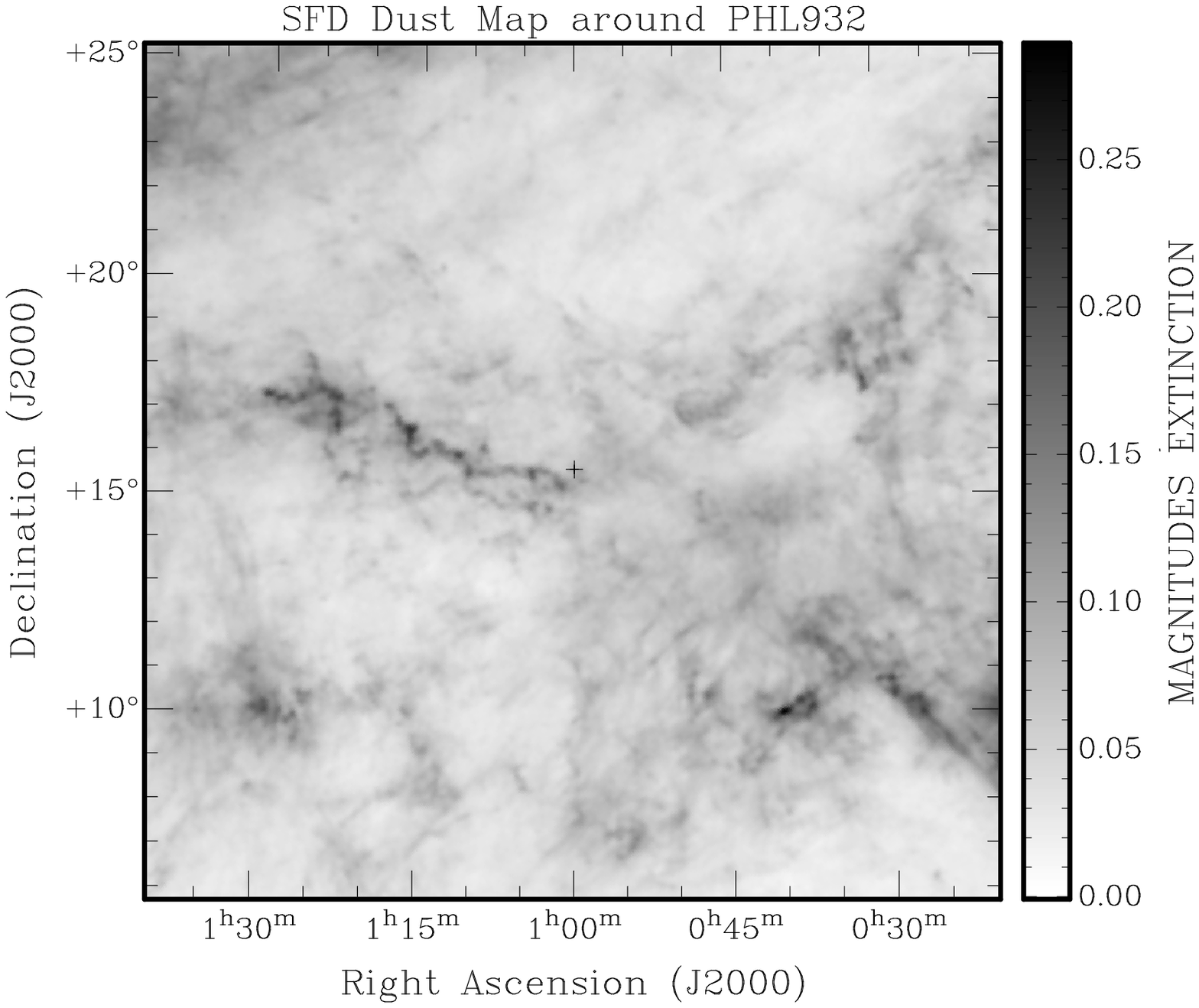}
\includegraphics[width=8.5cm]{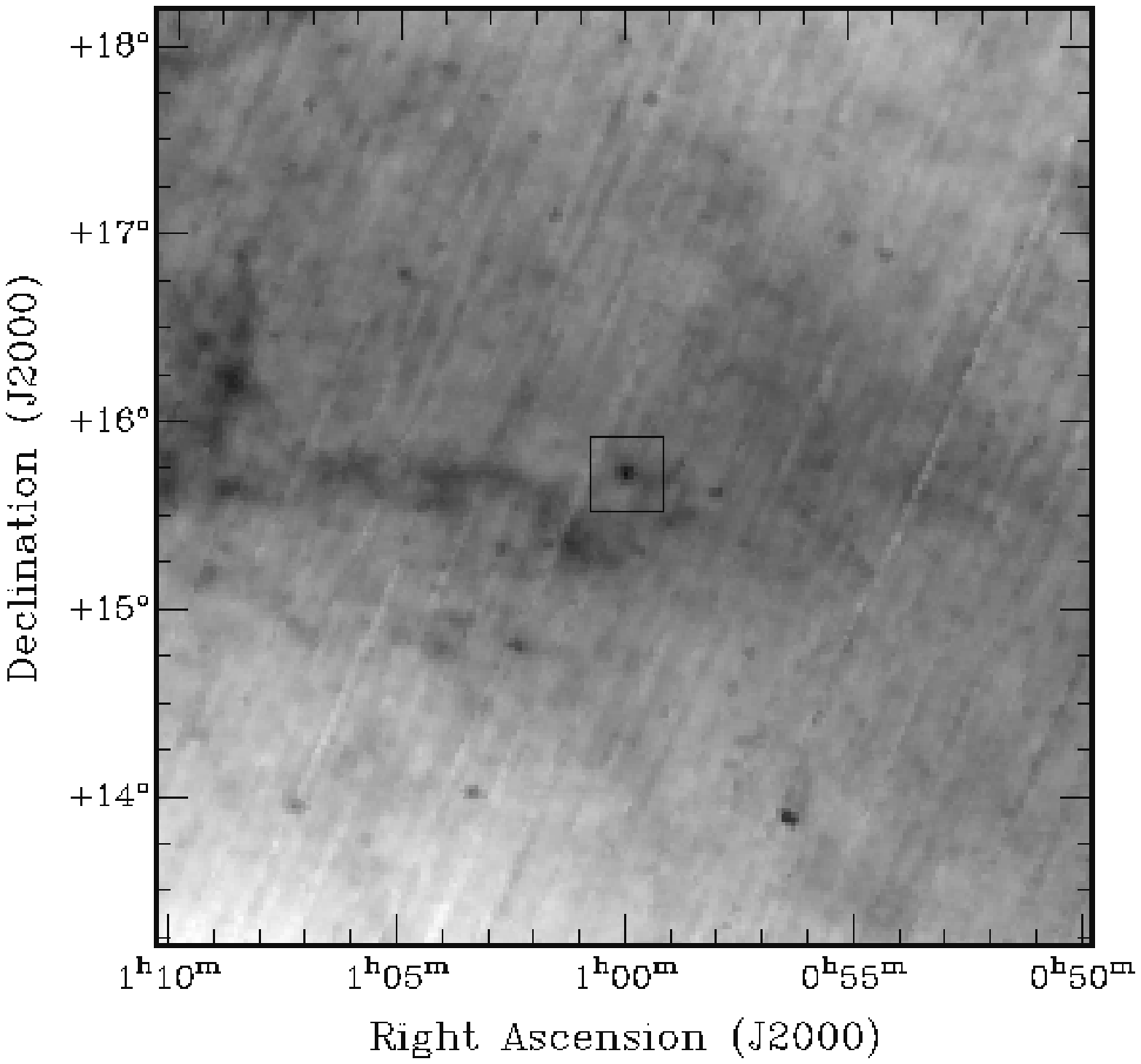}
\caption{Top panel: Region (20\arcdeg\ $\times$ 20\arcdeg) surrounding PHL~932 from the dust map of Schlegel et al. (1998).  The position of PHL~932 is marked with a cross and is found at the end of a filament of molecular emission associated with MBM~3 and MBM~4.  Lower panel:  IRAS 60~micron dust map centred on the position of PHL 932. The image is 5\arcdeg\ on a side.}
\label{fig:dust_map}
\end{center}
\end{figure}

It seems there is good evidence for widespread high-latitude gas and
molecular material in this general
direction, as seen in Figure~\ref{fig:dust_map}.  In Table~\ref{PHLenvironment}, 
we summarise the radial velocities
of the various  molecular clouds and emission nebulae within a
10\arcdeg\ radius of PHL~932.  The velocities for the molecular clouds
are taken from Magnani et al. (1985).   
In addition, Magnani et al. (2000) detected CO at $l,b$ = 127.4,
$-47.0$ on a sightline 1.5\arcdeg\ from the position of PHL~932.
They measured $V_{\rm LSR}$ = $-10.5$ kms$^{-1}$ and associated the
emission with the MBM~3 cloud.  The OH data for MBM~2 from Magnani \&
Siskind (1990) are also consistent in velocity.

As seen in Table~\ref{PHLenvironment}, all features have a mean
velocity near  $V_{\rm LSR}$ = $-8$ kms$^{-1}$, suggesting that the
widespread molecular and ionized gas in this direction is at a common
distance, $\sim$300 pc.  Crucially, the systemic velocity of the ionized gas around PHL~932 is identical within the errors with the mean velocity of the molecular gas in this direction.  We note, however, that Penprase (1992) has estimated an approximate distance to the MBM 3/4 complex of 90 pc $\leq$ $D$ $\leq$ 190 pc, so it is possible there is a considerable depth to the gas along this sightline.   

However, we ask if it is reasonable to expect interstellar gas with the observed emission measure at this location, $\sim$220 pc from the Galactic midplane?  It is widely known that there are large electron density variations in the ISM on all scales (e.g. Elmegreen 1998; Stinebring 2006) as seen in Figure~\ref{fig:dust_map}.  We also note that dense molecular gas is known to exist at remarkably large $|z|$ distances from the plane, as seen in the case of the Draco cloud which has $z$ \grtsim500 pc (Mebold et al. 1985; Penprase, Rhodes  \& Harris 2000).  Hence, we conclude that molecular gas and dust exists at moderate $z$ distances in the direction of PHL~932.

Finally, we comment on the emission nebulae associated with the hot DO white
dwarfs PG~0108+101 (Reynolds 1987; Frew, Parker \& Madsen 2006; Madsen
\& Frew in prep.) and PG~0109+111 (Reynolds 1987; Werner et al. 1997)
that are only $\sim$6\arcdeg\ southeast of PHL 932.  The fairly similar
apparent magnitudes, temperatures, and surface gravities of these two
stars (e.g. Wesemael, Green \& Liebert 1985) suggest they are at quite
similar distances; gravity distances range between 200 and 400~pc
(Pottasch 1996; Dreizler \& Werner 1996).  Considering the error bars
on the published log\,$g$ values, their distances are consistent both
with each other, and with the distance to PHL~932.  The systemic velocities of the gas are taken from unpublished WHAM data (Madsen \& Frew 2009, in preparation).  The consistency of these velocities with the mean velocity of the molecular gas in the area, as well as the narrow emission line widths from WHAM show that these nebulae are also Str\"omgren spheres in the ISM.  Their very low emission measures are consistent with the lack of obvious extinction in the SFD dust map (Figure~\ref{fig:dust_map}).

\begin{table*}	
{\footnotesize
\begin{center}
\caption{Summary of Radial Velocity Determinations.}
\bigskip
\label{PHLenvironment}
\begin{tabular}{lllllll}
\hline
Object            &  Type  & $l$  & $b$ & $V_{\rm LSR}$ & Line &  Ref\\
                     &            &       &        &      \tiny (kms$^{-1}$)       &     & \\						
\hline
MBM 2           &    MC        &   117.4    &   $-52.3$     &  $-7.3$         &CO & 1\\	
	---       &     ...                    &       ...         &         ...        &  $-7.4$         &OH & 2 \\
PHL 932         &    EN        &    125.9   &  $-47.1$     &    $-9.0$        & \ha, {[\rm N \sc ii]} &3\\	
PG 0108+101 &     EN       &  130.8     &   $-52.2$    &    $-8.6$        & \ha, {[\rm N \sc ii]}, {[\rm O \sc iii]} & 3\\
	---       &     ...                    &       ...         &         ...        &      $-11.0$                &\ha & 4\\
PG 0109+111 &     EN       &   131.1    &   $-51.2$     &    $-10.1$      & \ha  &3\\	
MBM 3           &  MC          &   131.3    &    $-45.7$    &     $-7.6$      &  CO &1\\	
MBM 4           &  MC          &    133.5   &  $-45.3$      &    $-8.7$       &  CO &1\\	
\hline
\end{tabular}
\end{center}
{\bf Notes}. LSR velocities are quoted for all emission nebulae (EN) and molecular clouds (MC) within 10\arcdeg\ of PHL~932, ordered by Galactic longitude.  MBM~3 and MBM~4 are associated with LBN bright nebulae (Lynds 1965).  References:  1.  Magnani et al. (1985); 2. Magnani \& Siskind (1990);  3. This work; 4. Reynolds (1987).
}
\end{table*}


\section{Conclusions}
\label{sec:disc}

We have used a combination of new and archival multi-wavelength observations to show that (a) PHL\,932 is unlikely to be in a close binary system as previously suggested, and (b) the emission nebulosity
surrounding PHL\,932 is \emph{not} a PN as
has been commonly assumed, nor is it physically associated with the star.  We conclude instead that PHL~932 is ionizing a ``clumpy'' region of the ISM as it travels through it; in other words, 
the emission nebula is simply a Str\"omgren zone (or HII region), rather than a PN.

We note that another putative PN, EGB~5 (Ellis, Grayson \&
Bond 1984) is also associated with an sdB star.   An unpublished MSSSO
2.3-m long-slit spectrum of the nebula taken by one of us (D.J.F)
shows a nebula of low excitation, similar to that around PHL~932.
Like PHL~932, the morphology of EGB 5 on DSS images militates
against it being a bona fide PN, though deep narrowband \ha\ images are required for a definitive conclusion.   
We likewise consider the nebula around EGB~5 to be another candidate
Str\"omgren zone in the ISM.   Such nebulae are seen around hot subdwarfs and white dwarfs when the ISM is sufficiently dense to give an emission measure that facilitates optical detection (cf. Tat \& Terzian 1999).  In future papers in this series, we will show that several other nearby emission nebulae currently assumed to be PN are also Str\"omgren zones in the ISM; see Frew \& Parker (2006) and Madsen et al. (2006) for preliminary results.

\section*{Acknowledgments}
D.J.F. gratefully acknowledges Macquarie University for a research scholarship, and also the Government of Western Australia for additional financial support.  This study used data from the Wisconsin H-Alpha Mapper (WHAM) and the Southern H$\alpha$ Sky Survey Atlas (SHASSA) which were produced with support from the National Science Foundation through grant AST-0607512.  This research has made use of the SIMBAD database, operated at the CDS, Strasbourg, France.  Data products from the Two Micron All Sky Survey (2MASS) were also utilised, which is a joint project of the University of Massachusetts and the Infrared Processing and Analysis Center/California Institute of Technology, funded by the National Aeronautics and Space Administration and the National Science Foundation.  We also thank the support staff at the Anglo-Australian Observatory for assistance in observing with SPIRAL, and the anonymous referee for a careful reading of the manuscript.

\end{document}